\documentclass[aps,twocolumn,pra,superscriptaddress,amsmath,amssymb]{revtex4-1} 
\usepackage{dcolumn}
\usepackage{bm}
\usepackage{amsmath}
\usepackage{txfonts}
\usepackage[T1]{fontenc}
\usepackage{xspace}
\usepackage{ulem}
\usepackage{comment}
\usepackage{braket}
\newcommand{\bras}[1]{\langle#1\rvert}
\newcommand{\kets}[1]{\lvert#1\rangle}

\newcommand{\means}[1]{\langle#1\rangle}

\newcommand{\chem}[1]{\mathrm{#1}}
\ifx\pdfoutput\undefined
\usepackage[dvipdfmx]{graphicx}
\usepackage[dvipdfmx]{hyperref}
\usepackage[dvipdfmx]{color}
\usepackage[dvipdfmx]{xcolor}
\else
\usepackage{graphicx}
\usepackage{hyperref}
\usepackage{color}
\usepackage{xcolor}
\usepackage[version=3]{mhchem}
\fi
\begin{document}
\let\emph\textit

\title{
Field-angle dependence of thermal Hall conductivity in magnetically ordered Kitaev-Heisenberg system 
  }
\author{Shinnosuke Koyama}
\author{Joji Nasu}
\affiliation{
  Department of Physics, Yokohama National University, Hodogaya, Yokohama 240-8501, Japan
}

\date{\today}
\begin{abstract}
We study magnetic excitations and thermal Hall effect on the Kitaev-Heisenberg model under magnetic fields.
By employing the spin-wave theory for the magnetic orders realized in this model, we examine the topological nature of the spin-wave dispersions and calculate the thermal Hall conductivity.
The comprehensive investigations on the field-angle dependence clarify that the thermal Hall conductivity is sensitive to the spin ordered pattern and excitation spectra of magnons;
this quantity is enhanced by the noncoplanar spin configurations and small magnon gap in the excitation spectrum.
On the other hand, we also find a common feature in the field-angle dependence of the thermal Hall conductivity.
It vanishes when the magnetic field is on the planes spanned by the spin axes.
We reveal that the behavior is intrinsic to the Kitaev-Heisenberg model in an applied field and demonstrate that the introduction of the off-diagonal spin interaction causes the disappearance of the feature in the thermal Hall conductivity.
\end{abstract}
\maketitle


\section{Introduction}

The Kitaev quantum spin model has been intensively investigated since the proposal by A.~Kitaev~\cite{kitaev,RevModPhys.87.1,Trebst2017pre,Hermanns2018rev,Knolle2019rev,takagi2019rev,Motome2020rev}.
In the model, the ground state is exactly shown to be a quantum spin liquid (QSL) with the fractionalization of spins into Majorana fermions, which are expected to be applicable to topological quantum computation~\cite{kitaev}.
The Kitaev model is thought to be realized in compounds with $4d$ or $5d$ transition metal ions with the strong spin-orbit coupling~\cite{Jackeli2009}.
In particular, the iridium oxides $A_2 {\rm IrO}_3$ ($A=$Li, Na) and $\alpha$-$\chem{RuCl_3}$ have been energetically studied as candidate materials of the Kitaev QSL, where the magnetism is governed by $j_{\rm eff}=1/2$ spins in $\chem{Ir^{4+}}$ or $\chem{Ru^{3+}}$ ions~\cite{prb90-041112,science356-1055,prl117-126403, nature15-733, cm29-493002, prb101-020414,prb91-144420,prb92-235119,Kubota2015}.
Although the QSL ground state should be realized in the Kitaev model, a magnetic order appears at the lowest temperature in the real materials.
To understand the origin of the magnetic ordering, one has introduced additional interactions such as the Heisenberg and off-diagonal $\Gamma$ terms in the Kitaev model~\cite{prb94-064435, prl109-187201, prl113-107201, prl105-027204, prl112-077204,prl110-097204}.
While a lot of efforts have been devoted to clarifying the global phase diagrams of the Kitaev-Heisenberg and Kitaev-Heisenberg-$\Gamma$ models~\cite{prb83-245104,khmodel,1/s, prr2-013014, prl110-097204, prr2-012021}, the relationship to the candidate materials, in particular, the realistic values of the exchange constants, has been still under debate~\cite{npj5-2}.

Recently, magnetic-field effects on the Kitaev QSL have attracted considerable attention.
It has been reported that, in the Kitaev candidate material $\alpha$-$\chem{RuCl_3}$~\cite{prb93-155143}, the magnetic field suppresses the zigzag order, and it disappears at $H_c\sim$7~T at the lowest temperature~\cite{prb91-180401, prl120-117204, prb96-041405}.
When the magnetic order disappears, a continuum in the magnetic excitation spectrum was observed above $H_c$ by the inelastic neutron scattering measurement~\cite{npj3-8}.
These results suggest that the magnetic field induces the Kitaev QSL.
To clarify the nature of the fractional quasiparticles in this state, the thermal transport has been measured in $\alpha$-$\chem{RuCl_3}$~\cite{Hirobe2017, prl120-217205,Yu2018, prl118-187203,prb99-085136, arxiv2102-11410}.
In particular, the half-quantized plateau has been observed in the thermal Hall conductivity~\cite{nature559-227, prb102-220404, arxiv2001-01899, arxiv2104-12184}, which offers convincing evidence for the emergence of Majorana fermions and the presence of a Majorana chiral edge mode.
Triggered by these experimental results, magnetic-field effects on the Kitaev-related systems have been studied theoretically~\cite{srep6-37925, prb98-060404, prb98-060405, nc10-530, nature10-2470, nature11-1639, prb101-100408, prl119-127204, prb98-060412, prr1-013014, prl126-147201, arxiv2004-06119, arxiv2104-02892, prb100-144445,prb101-060404}.
Moreover, classical phase diagrams of the Kitaev-Heisenberg and Kitaev-$\Gamma$ models under the magnetic field were obtained by the mean-field (MF) approach and Monte Carlo simulations~\cite{prr2-013014, 1/s, khmodel, prr2-012021, cm31-423002}.
In the phase diagrams, an applied magnetic field stabilizes various ordered states, including noncollinear and noncoplanar configurations.
Stimulated by the recent experimental results on the field-angle dependence of the thermal Hall conductivity~\cite{arxiv2001-01899}, the thermal Hall effect for various field directions has been examined in the Kitaev models with additional interactions in the Majorana fermion representation~\cite{nc10-530, arxiv2004-06119, prr1-013014} and spin-wave theory~\cite{prl126-147201, prb98-060404, prb98-060405, prr2-013014, arxiv2102-00014}.
However, details on the field-angle dependence of magnetic excitations and thermal Hall conductivity remain unclear in the Kitaev-related models.

In this paper, we investigate the magnetic-field effect on the transport properties of magnetically ordered states in the Kitaev-Heisenberg model on a honeycomb lattice.
By applying the MF approximation and linear spin-wave theory, we calculate the dispersion of magnons and the thermal Hall conductivity ascribed to the magnons by changing the direction and strength of an applied magnetic field.
We examine the magnetic phase diagram with respect to the field angles and reveal that noncoplanar magnetic orders tend to appear when the field is applied perpendicular to the honeycomb plane.
In this field direction, the thermal Hall conductivity is enhanced by noncoplanar spin configurations associated with the spin scalar chirality.
We also find that the enhancement can be induced by a small magnon gap in the dispersion relation.
On the other hand, the thermal Hall effect originating from the topological nature of the magnon dispersions is absent when the field is parallel to the bond of the honeycomb lattice. 
This behavior is indeed a consequence of the symmetry of the system.
Moreover, the field-angle dependence exhibits nodal lines where the thermal Hall conductivity vanishes.
The feature does not depend on the types of magnetic orders stabilized in the Kitaev-Heisenberg model.
We clarify that this is the intrinsic nature of the model and violated by introducing off-diagonal spin interactions such as the $\Gamma$ term.

This paper is organized as follows.
In the next section, we introduce the Kitaev-Heisenberg model in an applied magnetic field.
In Sec.~\ref{sec:method}, we present the method used in the present study.
The MF approximation and spin-wave theory are described in Secs.~\ref{sec:method_mf} and \ref{sec:sw_theory}, respectively.
Section~\ref{sec:thermal_hall} provides the formalism of the thermal Hall conductivity in the spin-wave theory.
The symmetry of the Kitaev-Heisenberg model is discussed in Sec.~\ref{sec:symmetry}.
The results are given in Sec.~\ref{sec:result}.
In Sec~\ref{sec:polarlized}, we show the field angle dependence of the thermal Hall conductivity in the spin-polarized state.
The results for the magnetic-field effect in the stripy and zigzag states are given in Secs.~\ref{sec::st} and \ref{sec:zigzag}.
In Sec.~\ref{sec:discussion}, we discuss the role of the scalar spin chirality in the thermal Hall effect and the effect of additional interactions.
Finally, Sec.~\ref{sec:summary} is devoted to the summary.

\section{Model}

\begin{figure}[t]
  \begin{center}
    \includegraphics[width=\columnwidth,clip]{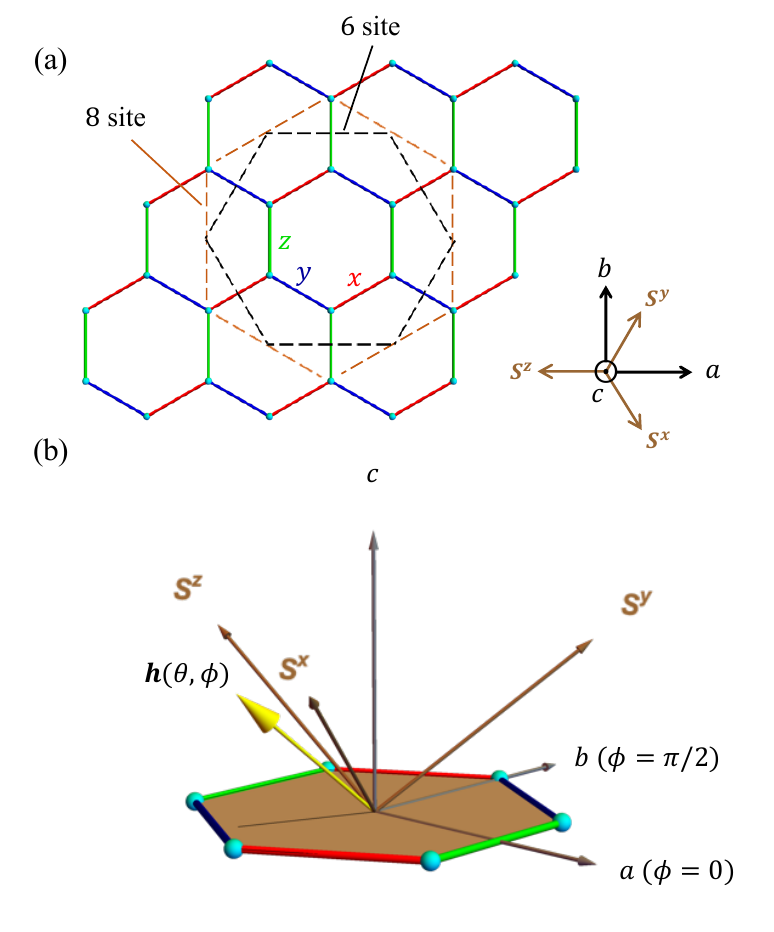}
    \caption{
      (a) Schematic picture of the honeycomb lattice on which the Kitaev-Heisenberg model is defined.
      The red, blue, and green lines stand for the $x$, $y$, and $z$ bonds, respectively.
      The dashed lines represent magnetic unit cells, including six and eight sites.
      The inset is the top view from the $c$ positive direction.
      (b) Spatial arrangement of the spin coordinate and the magnetic field, which is depicted by the yellow arrow.
      The magnetic-field angle is parametrized by the polar and azimuthal angles $(\theta,\phi)$ measured from the $c$ and $a$ axis, respectively.
    }
    \label{honey}
  \end{center}
\end{figure}

We consider the Kitaev-Heisenberg model on the honeycomb lattice shown in Fig.~\ref{honey}(a)~\cite{prl105-027204, prl112-077204, prl108-127203,prl110-097204}.
The model is given by
\begin{align}
  \label{kh}
  \mathcal{H}\ = \ J\sum_{\langle i,j \rangle} \bm{S}_i \cdot \bm{S}_j + 2K\sum_{\langle i,j \rangle_{\gamma}}  S_i^{\gamma} S_j^{\gamma}
  -\bm{h}\cdot \sum_{i}\bm{S}_i,
\end{align}
where $S_i^\gamma$ ($\gamma=x,y,z$) represents the $S=1/2$ spin at site $i$, and $K$ and $J$ are the exchange constants of the Kitaev and Heisenberg interactions, respectively, between spins on the nearest neighbor (NN) sites $\means{i,j}$.
In the Kitaev interaction, $\means{i,j}_\gamma$ denotes the NN $\gamma$ bond on the honeycomb lattice, which is depicted in Fig.~\ref{honey}(a). 
The exchange constants, $K$ and $J$, are parameterized by $\alpha \ (-\pi\leq \alpha \leq \pi )$ as $K = \sin\alpha$ and $J = \cos\alpha$.
The last term of Eq.~\eqref{kh} is the Zeeman term with the magnetic field $\bm{h}$.
Considering the correspondence to real materials, we introduce the spin coordinate such that the [111] direction in the spin space is parallel to the $c$ axis and the $S^z$ direction is on the $ac$ plane (see Fig.~\ref{honey}), where the $a$ axis is chosen so as to be perpendicular to the $z$ bonds of the honeycomb lattice.
In this situation, the $a$ and $b$ directions correspond to [$11\bar{2}$] and [$\bar{1}10$] in the spin coordinate~\cite{arxiv2001-01899,prl126-147201}.
The direction of the magnetic field is parameterized by the polar and azimuthal angles $(\theta,\phi)$ in the real space, which is depicted in Fig.~\ref{honey}(b).
Namely, the components $(h_a,h_b, h_c)$ of the magnetic field $\bm{h}$ along the $a$, $b$, and $c$ directions are given by
\begin{align}
  (h_a,h_b, h_c) = h(\sin\theta \cos\phi,\sin\theta \sin\phi, \cos\theta),
\end{align}
and the relationship to $(h_x, h_y, h_z)$ in the spin space is written as
\begin{align}
  \begin{pmatrix}
    h_x\\h_y\\ h_z
  \end{pmatrix}
  =\begin{pmatrix}
    1/\sqrt{6} & -1/\sqrt{2} & 1/\sqrt{3}\\
                    1/\sqrt{6} & 1/\sqrt{2}  & 1/\sqrt{3}\\
                    -2/\sqrt{6}& 0           & 1/\sqrt{3}
  \end{pmatrix}
  \begin{pmatrix}
    h_a\\h_b\\ h_c
  \end{pmatrix}.
\end{align}

In the absence of the magnetic field, the ferromagnetic (FM), N\'eel, zigzag, and stripy phases appear, although quantum fluctuations stabilize the QSL state near $\alpha=\pi/2$ and $-\pi/2$~\cite{prl110-097204}.
The classical phase diagrams for the Kitaev-Heisenberg model were obtained in the cases with $h \parallel S^z$ [$(\theta,\phi)=(\cos^{-1}(1/\sqrt{3}),\pi)$] and $\bm{h}\parallel c$ [$(\theta,\phi)=(0,0)$], which are qualitatively different from each other~\cite{khmodel}.
In the former case, the canted phases inherited from the N\'eel, zigzag, and stripy states appear at nonzero $h$ in addition to the polarized phase.
On the other hand, by applying the magnetic field along the $c$ axis, noncoplanar and multiple-$Q$ state states emerge additionally.
This is in contrast to the case with $h \parallel S^z$, where the coplanar spin configurations only appear.

\section{Method}
\label{sec:method}
\subsection{Mean-field theory}
\label{sec:method_mf}

In this section, we briefly introduce the MF theory for the localized spin model. 
We consider the quantum spin model, whose Hamiltonian is generally written as
\begin{align}
  \mathcal{H}\ = \ \sum_{\langle i,j \rangle} \sum_{\gamma \gamma'} \hat{J}^{\gamma\gamma'}_{ij} S_{i}^{\gamma} S_{j}^{\gamma'} - \sum_{i}\sum_{\gamma} h_{\gamma}S^{\gamma}_{i},
\end{align}
In the MF theory, the original Hamiltonian is decoupled to the two terms as
\begin{align}
  \mathcal{H} \ &= \ \mathcal{H}^{\chem{MF}} + \mathcal{H}^{\prime},\\
\end{align}
where the MF Hamiltonian is written as
\begin{align}
  \mathcal{H}^{\chem{MF}} \ &=\ \sum_{i} \mathcal{H}^{\chem{MF}}_{i} + \chem{const.},
\end{align}
and $\mathcal{H}^{\prime}$ is neglected in this approximation.
The local Hamiltonian $\mathcal{H}_i^{\chem{MF}}$ at site $i$ is given by
\begin{align}
  \mathcal{H}_{i}^{\chem{MF}} \ = \ \sum_{\gamma} \left( \sum_{\rho} \sum_{\gamma^{\prime}} \hat{J}^{\gamma\gamma^{\prime}}_{i\rho} \langle S^{\gamma^{\prime}} \rangle_{l_{\rho}} - h_{\gamma} \right) S^{\gamma}_{i}.
\end{align}
Here, $\rho$ stands for the nearest-neighbor sites of $i$, and $\langle S^\gamma \rangle_{l_i} = \braket{0;l_i|S^\gamma|0;l_i}$ represents the expectation value for the ground state $\ket{0;l_i}$ of the local Hamiltonian $\mathcal{H}^{\chem{MF}}_{i}$ where site $i$ belongs to the sublattice $l_{i}$.

In the present study, we perform the MF calculations in up to eight sublattices.
The unit cells including six and eight sites are depicted in Fig.~\ref{honey}(a).
Note that we cannot reproduce the two of the ordered phases for the present sublattice configurations: diluted star and zigzag star states, which appear in the classical phase diagram with $\bm{h}\parallel c$~\cite{khmodel}.
We avoid the parameters at which these states are stabilized.

\subsection{Linear spin-wave theory}
\label{sec:sw_theory}

To calculate the dispersion relation of elementary excitations from the magnetically ordered state determined by the MF approximation, we apply the linear spin-wave theory.
The contribution beyond the MF Hamiltonian, $\mathcal{H}^{\prime}$, is represented by 
\begin{align}
  \label{delh}
  \mathcal{H}^{\prime} \ &= \ \sum_{\langle ij \rangle}\sum_{\gamma\gamma^{\prime}}\hat{J}_{ij}^{\gamma \gamma^{\prime}} \delta S^{\gamma}_{i} \delta S_{j}^{\gamma^{\prime}},
\end{align}
where $\delta S^{\gamma}_{i}$ denotes the deviation from the MF value as follow:
\begin{align}
  \label{dels}
  \delta S_{i}^{\gamma}  \ &= \  S_{i}^{\gamma} - \langle S_{i}^{\gamma} \rangle _{l_{i}}.
\end{align}
Here, we apply the Holstein-Primakoff transformation to Eq.~\eqref{dels}~\cite{elementary,kusunose2001,prb103-L121104}.
By introducing the bosonic operator $a_i$ at each site $i$, we approximate $\delta S_{i}^{\gamma}$ as
\begin{align}
  \label{dels2}
  \delta S_{i}^{\gamma}\simeq \bras{1;l_i}\delta S^{\gamma} \kets{0;l_i}a_i^\dagger
  +\bras{0;l_i}\delta S^{\gamma} \kets{1;l_i}a_i,
\end{align}
where $\ket{1;l_{i}}$ is the excited state in $\mathcal{H}^{\chem{MF}}_{i}$.
Using this procedure, we approximately describe $\mathcal{H}^{\prime}$ as a bilinear form of the bosonic operators as $\mathcal{H}^{\prime}\simeq {\cal H}_{\rm SW}$, where ${\cal H}_{\rm SW}$ is given by
\begin{align}
  \label{hsw}
  \mathcal{H}_{\rm{SW}} \ &=\ \frac{1}{2}\sum_{\bm{k}} \mathcal{A}_{\bm{k}}^{\dagger} \mathcal{M}_{\bm{k}} \mathcal{A}_{\bm{k}},
\end{align}
and
\begin{align}
  \label{A}
  \mathcal{A}_{\bm{k}}^{\dagger} \ &= \ \left( a_{1,\bm{k}}^{\dagger}\ a_{2,\bm{k}}^{\dagger}\ \cdots \ a_{M,\bm{k}}^{\dagger} \ a_{1,-\bm{k}} \ a_{2,-\bm{k}} \ \cdots \ a_{M,-\bm{k}} \right).
\end{align}
Here, $M$ is the number of sublattices and $\mathcal{M}_{\bm{k}}$ is a $2M\times 2M$ Hermitian matrix.
The bosonic operator $a_{l,\bm{k}}$ is represented as
\begin{align}
  a_{l,\bm{k}}\ =\ \sqrt{\dfrac{M}{N}}\sum_{i\in l}a_{i} e^{-i\bm{k}\cdot \bm{r}_i},
\end{align}
where $N$ is total number of sites and $\bm{r}_i$ is the position of the lattice point in real space.
By applying the Bogoliubov transformation~\cite{colpa} to Eq.~\eqref{hsw}, the spin-wave Hamiltonian is represented as
\begin{align}
  \label{bogo}
  \mathcal{H}_{\rm{SW}} \ =\ \frac{1}{2}\sum_{\bm{k}} \mathcal{B}^{\dagger}_{\bm{k}} \mathcal{E}_{\bm{k}} \mathcal{B}_{\bm{k}},
\end{align}
where 
$\mathcal{B}_{\bm{k}}  = \mathcal{J}_{\bm{k}}^{-1} \mathcal{A}_{\bm{k}}$
and
$\mathcal{E}_{\bm{k}}=\mathcal{J}^{\dagger}_{\bm{k}}{\cal M}_{\bm{k}}\mathcal{J}_{\bm{k}}$.
The $2M\times 2M$ matrix $\mathcal{J}_{\bm{k}}$ is the paraunitary, which satisfies the relation $\mathcal{J}_{\bm{k}} \sigma_3 \mathcal{J}_{\bm{k}}^{\dagger} =\mathcal{J}_{\bm{k}}^{\dagger} \sigma_3 \mathcal{J}_{\bm{k}} = \sigma_3$ with the paraunit matrix $\sigma_3 = \left(\begin{smallmatrix} \bm{1}_{M\times M} & 0\\ 0& -\bm{1}_{M\times M} \end{smallmatrix} \right) $, where $\bm{1}_{M\times M}$ is the $M\times M$ unit matrix.
$\mathcal{E}_{\bm{k}}$ is the diagonal matrix given by
$\mathcal{E}_{\bm{k}} = {\rm diag}\{\varepsilon_{1,\bm{k}}, \varepsilon_{2,\bm{k}}, \cdots, \varepsilon_{M,\bm{k}},\varepsilon_{1,-\bm{k}}, \varepsilon_{2,-\bm{k}}, \cdots, \varepsilon_{M,-\bm{k}}\}$.
where $\varepsilon_{n\bm{k}}$ is the excitation energy for the $n$-th branch.

\begin{figure*}[t]
  \begin{center}
    \includegraphics[width=180mm,clip]{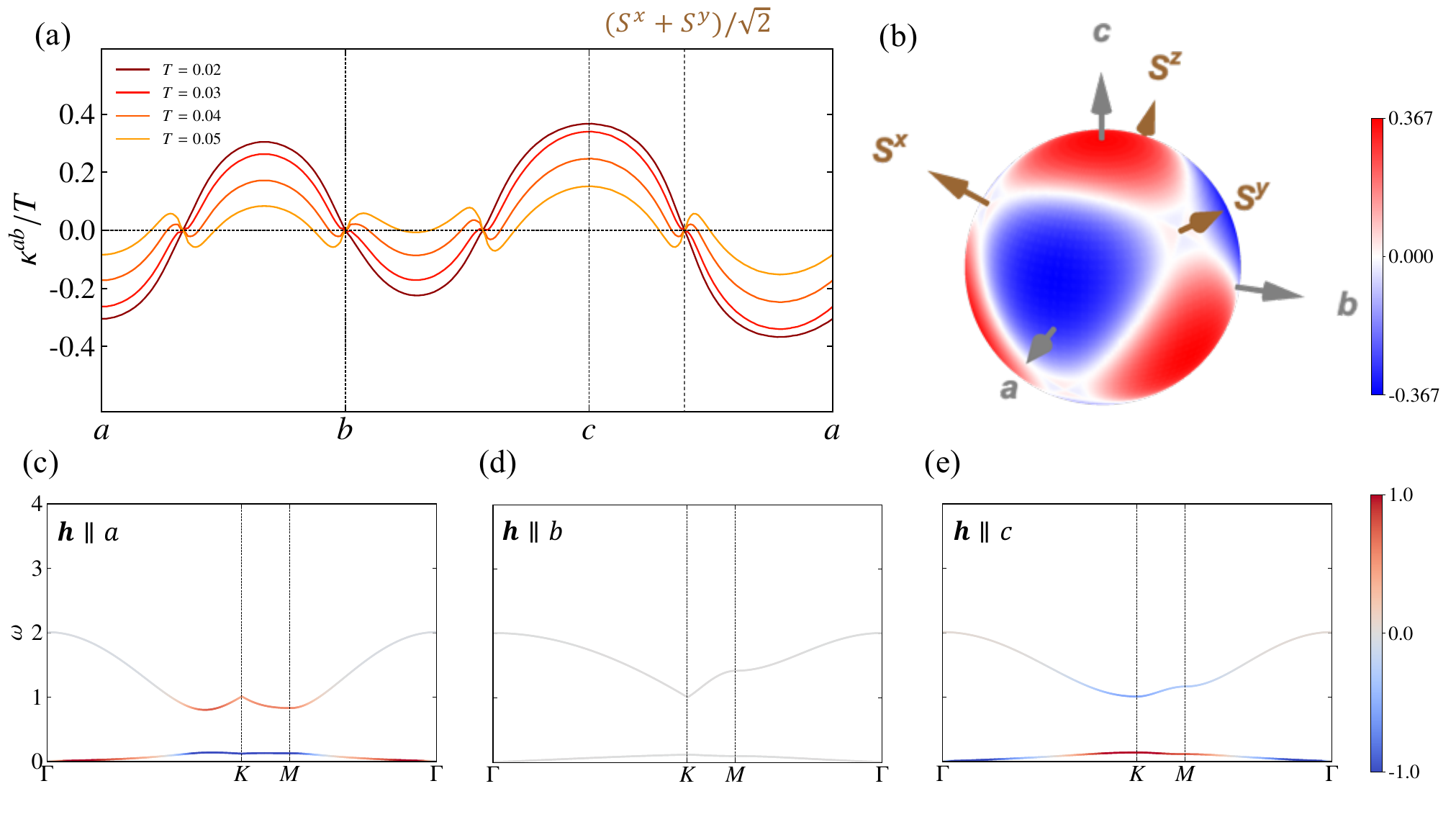}
    \caption{
(a) Thermal Hall conductivity $\kappa^{ab}/T$ as a function of the field angle in the FM Kitaev model ($\alpha=-\pi/2$) for the fixed field strength $h=0.01$ at several temperatures, and (b) its spherical plot at $T=0.05$.
The gray and orange arrows represent the real-space axes and the spin axes, respectively.
(c)--(e) Spin-wave dispersions under the magnetic fields applied along (c) the $a$, (d) $b$ and (e) $c$ directions.
The parameters used here are the same as those in (b).
The line color indicates the Berry curvature $\Omega_{n\bm{k}}$ associated with the magnon branch $\varepsilon_{n\bm{k}}$.
    }
    \label{ffmth}
  \end{center}
\end{figure*}

\subsection{Thermal Hall conductivity}
\label{sec:thermal_hall}

In this section, we introduce the expression of the thermal Hall conductivity $\kappa^{ab}$ in the spin-wave approximation~\cite{matsumoto2014, murakami2017, prl104-066403, prl103-047203}.
The thermal Hall conductivity is defined as
$\means{J_Q^a}_{\nabla T} = \kappa^{ab} (-\nabla _b T)$, where $\nabla _b T$ is the thermal gradient along the $b$ direction and $\means{J_Q^a}_{\nabla T}$ is the expectation value of the heat current along the $a$ axis in the presence of the thermal gradient.
The heat current is introduced as 
\begin{align}
  \bm{J}_Q=\frac{\partial \bm{P}_E}{\partial t}=\frac{i}{\hbar}[{\cal H},\bm{P}_E],\label{eq:JQ}
 \end{align}
where $\bm{P}_E$ is the energy polarization given by
\begin{align}
  \bm{P}_E=\sum_{i}\bm{r}_i h_i.
 \end{align}
The local Hamiltonian $h_i$ is defined from the terms of the Hamiltonian involving site $i$ so as to satisfy ${\cal H}=\sum_{i}h_i$.
The effect of the thermal gradient is introduced by replacing the Hamiltonian to $(1+\phi_g){\cal H}$, where $\phi_g$ is the pseudogravitational potential~\cite{matsumoto2014,murakami2017}.
Evaluating the linear response with respect to $\phi_g$, one can calculate $\kappa^{ab}$ as 
\begin{align}
  \label{th}
\kappa^{ab} = -\dfrac{k_B^2T}{\hbar V}\sum_{\bm{k}}^{\rm{BZ}}\sum_{n=1}^{M} c_2(f_{\rm{BE}}(\varepsilon_{n\bm{k}}))\Omega_{n\bm{k}},
\end{align}
where  $c_2(x)$ is defined as
\begin{align}
  \label{c2}
  c_2(x) = \int_0^{x} dt \left( \ln \dfrac{1+t}{t} \right)^2,
\end{align}
and $f_{\rm{BE}}(\varepsilon_{n\bm{k}})$ is the Bose distribution function, and BZ denotes the Brillouin zone for the magnetic unit cell.
Note that $c_2(f_{\rm BE}(\varepsilon))$ is positive and quickly decays from $\pi^2/3$ to zero while increasing $\varepsilon$ at low temperatures.
$\Omega_{n\bm{k}}$ is the Berry curvature in the momentum space, which is defined as
\begin{align}
  \label{bc1}
  \Omega_{n\bm{k}}= i\left[ \sigma_3 \dfrac{\partial \mathcal{J}_{\bm{k}}^{\dagger}}{\partial k_a} \sigma_3 \dfrac{\partial \mathcal{J}_{\bm{k}}}{\partial k_b}  \right]_{nn} + {\rm H.c.}
\end{align}
In this equation, we need to evaluate the momentum derivatives of $\mathcal{J}_{\bm{k}}$, which are difficult to be obtained in the numerical calculations.
To avoid the computation of the momentum derivatives, we rewrite Eq.~\eqref{bc1} to
\begin{align}
  \label{bc2}
  \Omega_{n\bm{k}} = -2\sum_{m(\neq n)}^{2M}{\rm Im}\left[\dfrac{ (\sigma_3)_n (\sigma_3)_m \left( \mathcal{J}^{\dagger}_{\bm{k}} v^a \mathcal{J}_{\bm{k}} \right)_{nm} \left( \mathcal{J}^{\dagger}_{\bm{k}} v^b \mathcal{J}_{\bm{k}} \right)_{mn} }{ \Big\{ ( \sigma_3 \mathcal{E}_{\bm{k}} )_{m} - ( \sigma_3 \mathcal{E}_{\bm{k}} )_n \Big\}^2 }\right],
\end{align}
where $v^{\mu} = \partial \mathcal{M}_{\bm{k}}/\partial k_{\mu}$~\cite{prb99-014427}. 
If a magnon branch is well separated from others, one can define the Chern number $N_n^{\rm Ch}$ for the corresponding branch $n$ as \cite{prb87-174427}
\begin{align}
  \label{ch}
  N_n^{\rm Ch}= \int_{\chem{BZ}} \frac{dk_x\ dk_y}{2\pi} \Omega_{n\bm{k}},
\end{align}

\subsection{Symmetry argument}
\label{sec:symmetry}

Before showing the results for the numerical calculations, we discuss the properties of the thermal Hall conductivity based on the symmetry of the Hamiltonian and lattice structure.
There are two symmetric operations in the present system: the $C_3$ rotation around the $c$ axis and $C_2$ rotation around $b$ axis or its equivalent directions.
In the former, the spin axes are changed cyclically, such as $(S^x,S^y,S^z)\to (S^y,S^z,S^x)$.
On the other hand, the spin coordinate $(S^x,S^y,S^z)$ is transformed to $(-S^y,-S^x,-S^z)$ by the latter operation. 
In the absence of the magnetic field, the Kitaev-Heisenberg model given in Eq.~\eqref{kh} is invariant under these operations.
Moreover, the invariance is retained even with the $\Gamma$ and $\Gamma'$ interactions~\cite{arxiv2004-06119}.
With respect to the thermal Hall conductivity, $\kappa^{ab}$ should be unchanged by the $C_3$ rotation around the $c$ axis, but it changes its sign under the $C_2$ rotation around the $b$ axis.
This is because the latter operation flips the $a$ direction, and thereby, $\kappa^{ab}$ should change to $\kappa^{(-a)b}=-\kappa^{ab}$.
If the magnetic field is applied along the $b$ direction, the Hamiltonian is unchanged under the $C_2$ rotation around $b$ axis.
Therefore, $\kappa^{ab}$ should be zero in the presence of the magnetic field parallel to $b$ or its equivalent directions when a magnetic order does not occur~\cite{arxiv2001-01899,arxiv2004-06119, prb103-155124}.

In addition to this symmetry, there is an additional property specific to the Kitaev-Heisenberg model.
In the absence of the magnetic field, this model is trivially written by a real symmetric matrix when choosing the spin bases appropriately.
The feature is maintained even in the presence of the magnetic field applied on the plane of the $S^x$-$S^z$ plane.
On the other hand, Eq.~\eqref{eq:JQ} indicates that the heat current is represented by the Hermitian matrix with pure imaginary matrix elements as the Hamiltonian is real.
Even in the presence of the pseudogravitational potential, the Hamiltonian is also real, and hence, $\means{J_Q^a}_{\nabla T}=0$.
Thus, $\kappa^{ab}$ vanishes when the magnetic field is on the $S^xS^y$, $S^yS^z$, and $S^zS^x$ planes (the planes involving two of the spin axes).
We refer to them as the spin axis planes hereafter.
In this case, $h_x h_y h_z=0$ is satisfied, and the above condition includes the case of $\bm{h}\parallel b$.
We note that Kitaev pointed out that $\kappa^{ab}/T$ is quantized as ${\rm sgn}(h_x h_y h_z) \times \pi/12$ in the Majorana fermion picture when the magnetic field is small enough~\cite{kitaev}.
This is consistent with the above condition.

When a magnetic order occurs, one should discuss the properties of ${\cal H}'$ given in Eq.~\eqref{delh} instead of ${\cal H}$.
Suppose the magnetic order is coplanar and the plane is parallel to the applied magnetic field. In that case, the above discussion can be applied to ${\cal H}'$ as it is written as a real matrix on appropriate spin bases.
In the spin-wave approximation, this corresponds to the fact that the coefficients of the bosonic operators in 
Eq.~\eqref{dels2} is real.

The properties of $\kappa^{ab}$ vanishing in the case of $h_x h_y h_z=0$ is violated by the introduction of off-diagonal spin interactions such as the $\Gamma$ term.
The properties in the presence of the $\Gamma$ interaction will be discussed in Sec.~\ref{sec:discussion}.
The nonzero $\kappa^{ab}$ in a magnetic field on the spin axis planes is understood as the contribution beyond the Kitaev-Heisenberg model.

\section{Result}
\label{sec:result}

\subsection{Spin-polarized state}
\label{sec:polarlized}

\begin{figure}[t]
  \begin{center}
    \includegraphics[width=\columnwidth,clip]{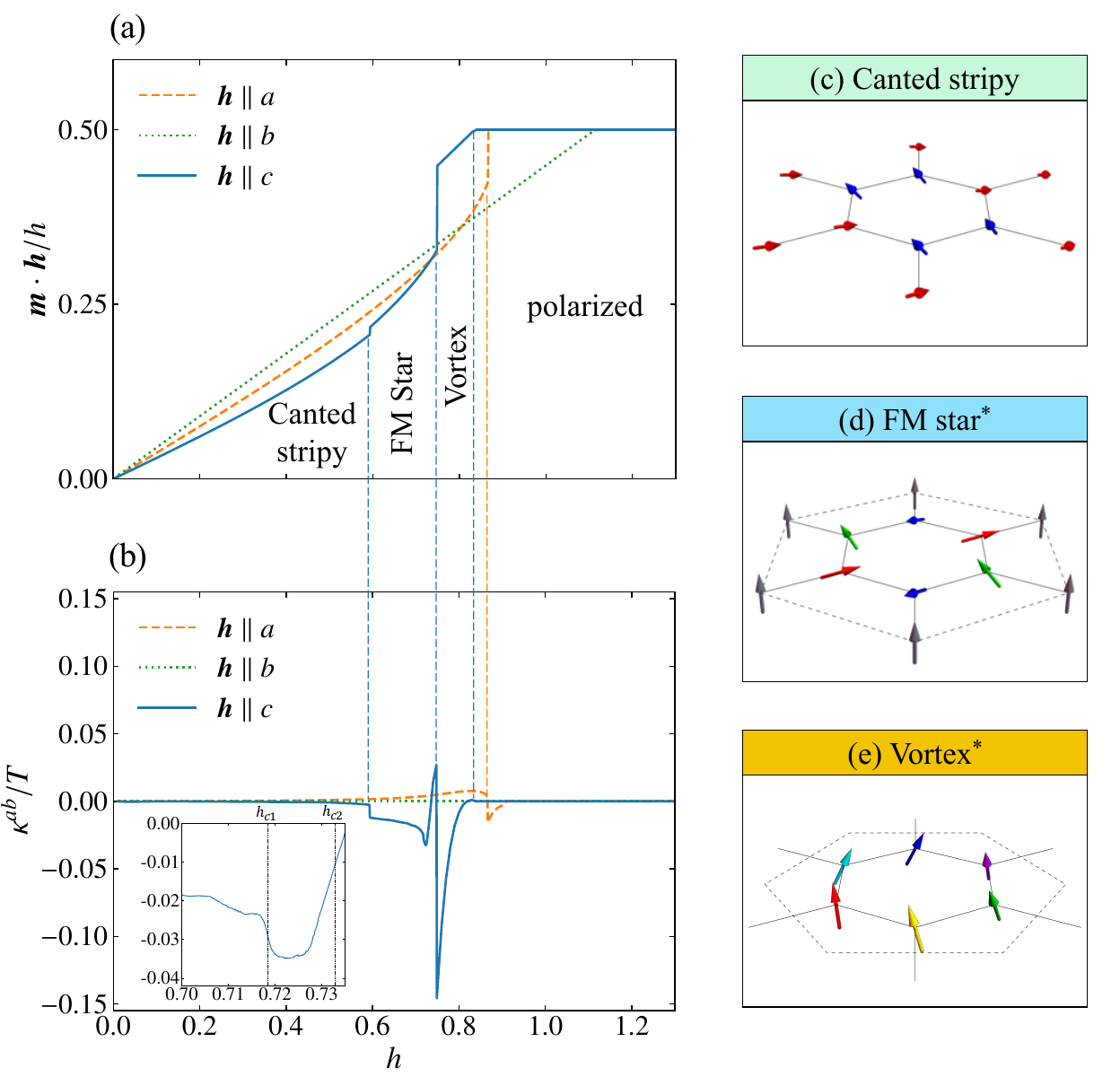}
    \caption{
(a) Total magnetization $\bm{m}$ and (b) thermal Hall conductivity $\kappa^{ab}/T$ of the Kitaev-Heisenberg model with $\alpha = -0.312\pi$ in the magnetic field along the $a$, $b$, and $c$ directions as a function of field strength $h$.
The inset of (b) shows the extended plot in the FM star phase.
(c)--(e) The spin configurations of (c) the canted stripy, (d) FM star, and (e) vortex states.
Arrows with the same color denote the equivalent spins.
The FM star and vortex states are eight-sublattice and six-sublattice spin configurations surrounded by the dotted lines and exhibit non-coplanar spin structures, indicated by asterisks.
  }
  \label{stmag}
  \end{center}
\end{figure}
\begin{figure}[t]
  \begin{center}
    \includegraphics[width=\columnwidth,clip]{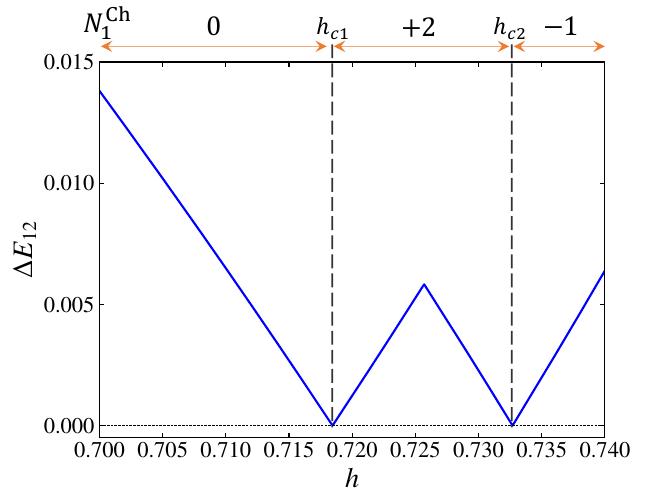}
    \caption{
      Energy gap between the lowest and second-lowest magnon branches in the FM star phase in the Kitaev-Heisenberg model with $\alpha = -0.312\pi$ for $\bm{h}\parallel c$.
      The Chern number of the lowest energy branch is shown at the top of the figure.
  }
  \label{fmhd}
  \end{center}
\end{figure}

\begin{figure*}[t]
  \begin{center}
    \includegraphics[width=180mm,clip]{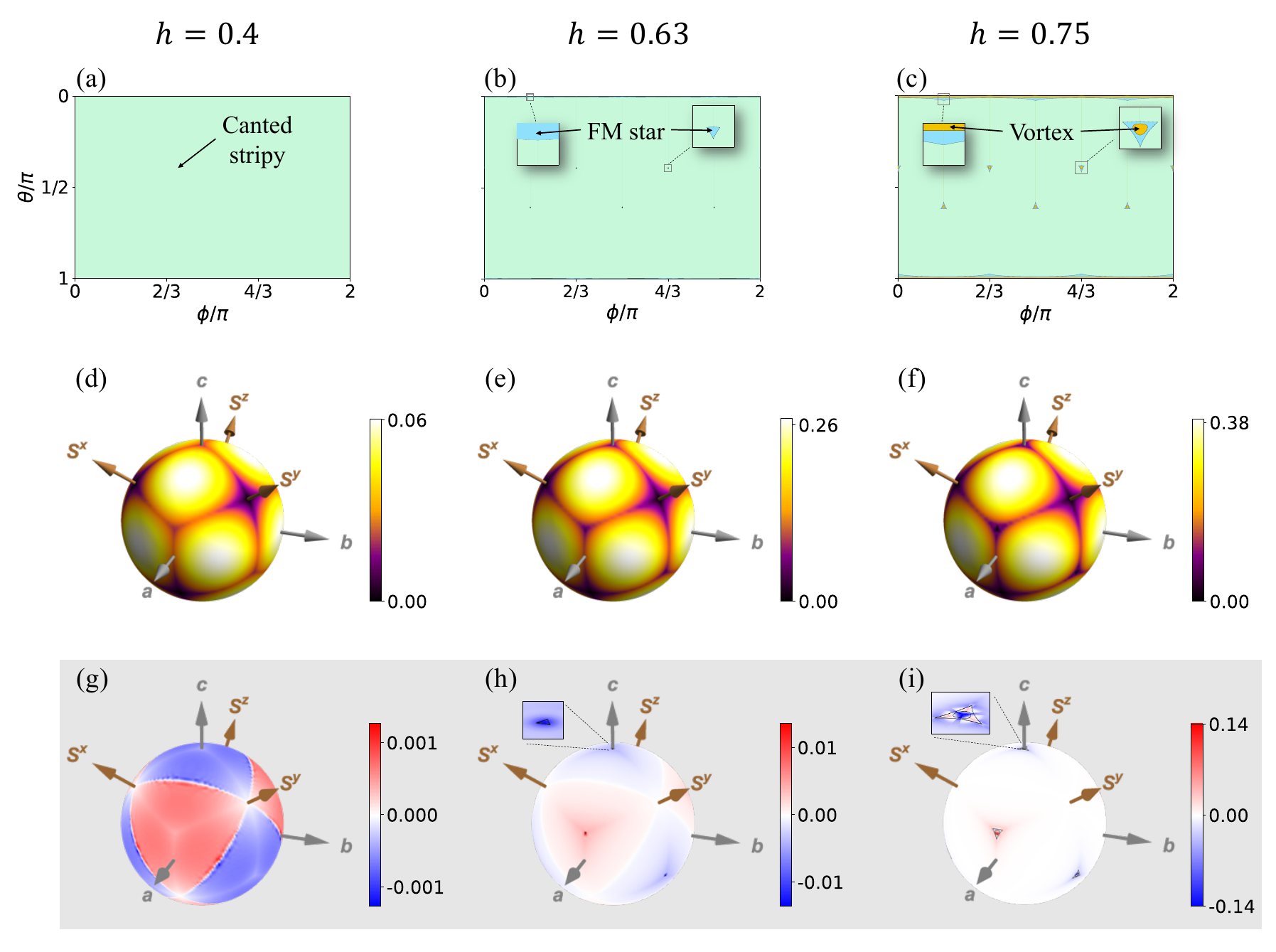}
    \caption{
 (a) Classical phase diagram of the Kitaev-Heisenberg model with $\alpha = -0.312\pi$ on the plane of the field angles $(\theta,\phi)$ in the field strength fixed to $h=0.4$.
 (d),(g) Spherical plots of (d) the magnon gap and (g) $\kappa^{ab}/T$ at $(T,\alpha)=(0.05,-0.312\pi)$ and $h = 0.4$.
(b),(e),(h) and (c),(f),(i) Corresponding plots at $h=0.63$ and $h=0.75$, respectively.
   }
   \label{stresult}
   \end{center}
 \end{figure*}
\twocolumngrid

First, we show the results for the spin-polarized phase at $(\alpha,\ h) = (-\pi/2,\ 0.01) $, which corresponds to the FM Kitaev model under the magnetic field.
Figure~\ref{ffmth}(a) shows the field-angle dependence of $\kappa^{ab}/T$ at several temperatures.
We set $k_B$, $\hbar$, and the length of the primitive translation vectors of the honeycomb lattice to be unity.
With increasing temperature, the magnitude of $\kappa^{ab}/T$ becomes small for almost all the field angles.
We find that $\kappa^{ab}$ is zero for $\bm{h}\parallel b$ while it is nonzero for $\bm{h}\parallel a,c$.
The features are understood from the symmetry, which was discussed in Sec.~\ref{sec:symmetry}.
Moreover, in addition to $h\parallel b$, the field angles vanishing $\kappa^{ab}$ are found in the spin-polarized phase.
For example, $\kappa^{ab}=0$ when the magnetic field is parallel to $(S^x+S^y)/\sqrt{2}$ as shown in Fig.~\ref{ffmth}(a).
To discuss the issue more detail, we show the spherical plot of $\kappa^{ab}/T$ at $T=0.02$ in Fig.~\ref{ffmth}(b).
As presented in this figure, this is zero when the magnetic field is on the spin axis planes, which include the direction along $(S^x+S^y)/\sqrt{2}$.
The result is consistent with the discussion described in Sec.~\ref{sec:symmetry}.

The sign of $\kappa^{ab}/T$ is dependent on the angle of the magnetic field.
As shown in Figs.~\ref{ffmth}(a) and \ref{ffmth}(b), this is negative (positive) for $\bm{h}\parallel a$ ($\bm{h}\parallel c$).
To clarify the origin of the sign, we calculate the Berry curvature given in Eq.~\eqref{bc1}.
Figures~\ref{ffmth}(c)--\ref{ffmth}(e) show the momentum dependence of $\Omega_{n\bm{k}}$ in the branch of $\omega=\varepsilon_{n\bm{k}}$ for the cases with $\bm{h}\parallel a$, $b$, and $c$, respectively.
For $\bm{h}\parallel b$, the Berry curvature is always zero for the all branches, leading to the zero $\kappa^{ab}$.
On the other hand, the Berry curvature is nonzero and exhibits the momentum dependence for the other directions.
In the case of $\bm{h}\parallel c$, the Chern numbers for the lower and upper bands are $+1$ and $-1$, respectively.
As shown in Fig.~\ref{ffmth}(e), the sign of $\Omega_{n\bm{k}}$ depends on $\bm{k}$ and the low-energy part of $\Omega_{n\bm{k}}$ is negative although its momentum integral is positive.
Since the negative sign is present in Eq.~\eqref{th}, $\kappa^{ab}$ is positive.
We have confirmed that the Chern number of each branch is not changed while increasing $h$, and the sign of $\kappa^{ab}$ is inverted to negative by temperature because of the positive Chern number in the lower band.
The results are consistent with the previous study~\cite{prb98-060404}.
In the case of $\bm{h}\parallel a$ [Fig.~\ref{ffmth}(c)], the sign of $\Omega_{n\bm{k}}$ is opposite to that in $\bm{h}\parallel c$.
This leads to the negative value of $\kappa^{ab}$.

\subsection{Canted stripy, FM star, and vortex states}
\label{sec::st}

In this section, we show the result for $\alpha = -0.312\pi$, where the stripy state is stabilized in the absence of the magnetic field~\cite{prl110-097204}.
Figure~\ref{stmag}(a) shows the total magnetization $\bm{m}$ projected onto the field direction in the magnetic fields along $a$, $b$, and $c$ directions, where $\bm{m}=N^{-1}\sum_i\bm{S}_i$.
In the case of $\bm{h}\parallel a$, the canted stripy state shown in Fig.~\ref{stmag}(c) appears by introducing the magnetic field.
Eventually, the fully spin-polarized state along the field direction is stabilized above $h\gtrsim 0.867$ with the magnetization jump.
A similar behavior is observed for $\bm{h}\parallel b$ but the magnetization is continuous at the boundary between the canted stripy and spin-polarized phases.
In this case, the magnetization linearly increases with increasing $h$ in the canted stripy phase while it is not linear for $\bm{h}\parallel a$.
The difference originates from the fact that $\bm{m}$ is parallel to the field direction for $\bm{h}\parallel b$ but is not for $\bm{h}\parallel a$, where the angle between $\bm{m}$ and $\bm{h}$ depend on the strength of the applied field~\cite{khmodel}.

Although the canted stripy and spin-polarized phases are stabilized for $\bm{h}\parallel a$ and $b$, the magnetic field parallel to the $c$ direction yields the additional states, the FM star [Fig.~\ref{stmag}(d)] and vortex states [Fig.~\ref{stmag}(e)], in between these phases~\cite{khmodel}.
Note that the total magnetization for the FM star and vortex states is parallel to the field direction but is not in the canted stripy state.
This results in a continuous change of the magnetization at the boundary between the vortex and polarized states~\cite{khmodel}.
We note that the FM star and vortex states possess noncoplanar spin configurations accompanied by a magnon gap.

Next, we examine the transport properties and discuss the relationship to such peculiar spin configurations.
Figure~\ref{stmag}(b) shows the thermal Hall conductivity as a function of the field strength at $T=0.05$.
For the case with $\bm{h}\parallel b$, $\kappa^{ab}$ is always zero independent of the field strength.
This result originates from the fact that the Berry curvature $\Omega_{n\bm{k}}$ is zero, similar to the case for the spin-polarized state [Fig.~\ref{ffmth}(d)].
On the other hand, for $\bm{h}\parallel a$, $\kappa^{ab}$ is nonzero.
In particular, the absolute value of $\kappa^{ab}$ is enhanced around the phase boundary between the canted stripy and spin-polarized phases while its sign is inverted between these phases.
While changing the field strength away from the phase boundary, the absolute value of $\kappa^{ab}$ gradually decreases and goes to zero.

The thermal Hall conductivity is nonzero also for the magnetic field parallel to the $c$ direction.
In the canted stripy phase, $|\kappa^{ab}|$ gradually increases with increasing $h$.
When the field strength is beyond the boundary with the FM star state, this quantity is strongly enhanced.
The absolute value of the thermal Hall conductivity in the FM star and vortex states is substantially larger than that in the canted stripy phase.
We expect that the noncoplanar spin configurations are relevant to the topological nature of the magnon bands and play a crucial role in the enhancement of the thermal Hall conductivity.
We also find that $\kappa^{ab}$ is continuously changed at the phase boundary between the vortex and spin-polarized states because of the continuous change of the total magnetization, unlike at the phase boundary between the canted stripy and FM star states.

As shown in the inset of Fig.~\ref{stmag}(b), in the FM star phase for $\bm{h}\parallel c$, we find the nonmonotonic $h$ dependence of $\kappa^{ab}$.
To clarify the origin, we calculate the energy difference between the lowest and second-lowest magnon branches, $\Delta E_{12}$, and Chern number of the lowest magnon branch, $N_1^{\rm Ch}$.
As shown in Fig.~\ref{fmhd}, we find that $\Delta E_{12}$ vanishes at $h_{1c}\simeq 0.7184$ and $h_{2c}\simeq 0.7327$ and $N_1^{\rm Ch}$ is nonzero above $h_{c1}$.
Moreover, $N_1^{\rm Ch}$ changes from $2$ to $-1$ at $h_{2c}$ while increasing $h$.
The change of the Chern number causes the decrease of $\kappa^{ab}$ around $h_{c1}$ and the increase around $h_{c2}$ because of the negative sign in Eq.~\eqref{th}.
Note that, below $h_{c1}$, the Berry curvature is nonzero although $N_{1}^{\rm Ch}=0$, which causes the nonzero $\kappa^{ab}$.

\begin{figure}[t]
  \begin{center}
    \includegraphics[width=\columnwidth,clip]{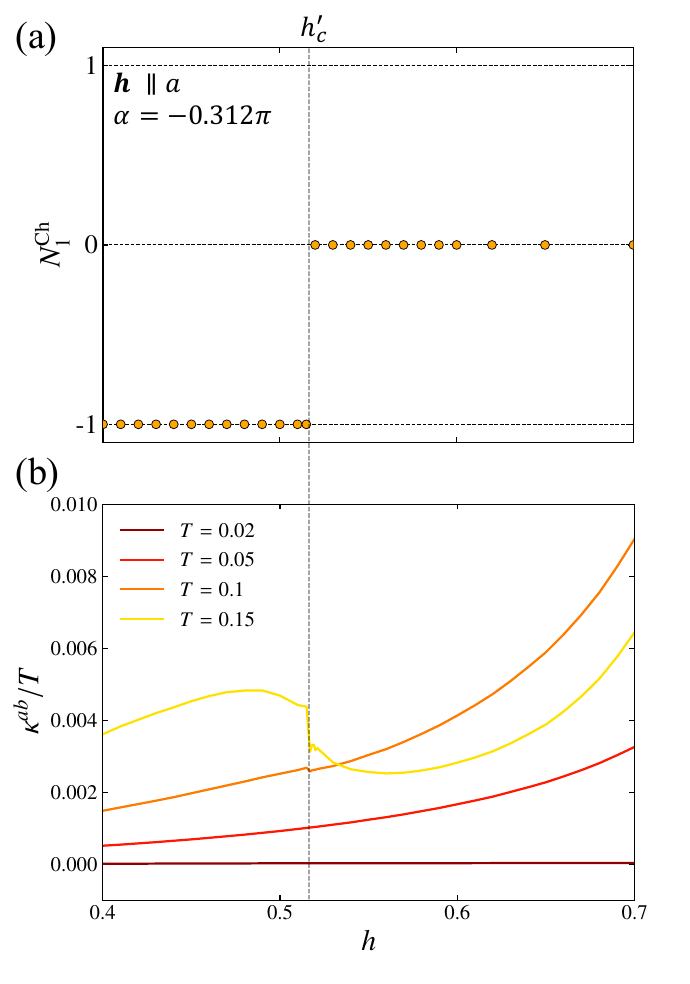}
    \caption{
      (a) Field-strength dependence of the Chern number of the lowest magnon branch and (b) the thermal Hall conductivity in the Kitaev-Heisenberg model with $\alpha = -0.312\pi$ under the magnetic field along the $a$ direction.
    }
    \label{fig:ach}
    \end{center}
  \end{figure}

While there appears to be no anomaly in the $h$ dependence of $\kappa^{ab}$ for $\bm{h}\parallel a$, we find a topological transition in the canted stripy phase.
Figure~\ref{fig:ach}(a) shows $N_1^{\rm Ch}$ as a function of the strength of the magnetic field parallel to the $a$ direction.
A topological transition from $-1$ to $0$ is found at $h'_{c}\simeq 0.52$.
Note that the lowest magnon branch is well separated from the others except for the transition point.
Figure~\ref{fig:ach}(b) shows $\kappa^{ab} /T$ for several temperatures.
Although we do not find any anomaly in $\kappa^{ab}$ at $T=0.02$ and $0.05$, the abrupt decrease at $h'_c$ are exhibited at $T=0.1$ and $T=0.15$, which is a consequence of the topological transition associated with the increase of the Chern number $N_1^{\rm Ch}$ for the lowest energy branch.

Here, we discuss the field-angle dependence of the excitation gap of magnons and the thermal Hall effect for the fixed field strength in the model with $\alpha=-0.312\pi$.
Figure~\ref{stresult}(a) shows the ground state phase diagram on the plane of the field angles $(\theta,\phi)$ at $h=0.4$.
In this case, the canted stripy state is realized regardless of the field direction.
With increasing $h$, additional phases appear.
Figures~\ref{stresult}(b) and \ref{stresult}(c) show the phase diagrams at $h=0.63$ and $0.75$, respectively.
The FM star state is stabilized around the $c$ axis at $h=0.63$, and further increase of the field strength stabilizes the vortex state, which is surrounded by the FM star phase at $h=0.75$.
We note that the magnon gap is present for the canted stripy and FM star phases, but the vortex state possesses a gapless magnon mode.

While the spin configuration of the ground state is not altered by changing the field direction for the weak field strength, the magnon gap strongly depends on it.
Figure~\ref{stresult}(d) shows the spherical plot of the magnon gap at $h=0.4$.
We find that the gap vanishes when the magnetic field is parallel to the spin axes.
Moreover, the gap takes minimal values on the lines from these spin axes to the $c$ axis.
Such lines divide the sphere into twelve, on which the field component taking the smallest value among $h_x$, $h_y$, and $h_z$, is switched.
With increasing $h$, the magnon gap increases, but the overall structure of the angle dependence is almost unchanged as shown in Figs~\ref{stresult}(e) and \ref{stresult}(f).
Note that the gap is always zero in the magnetic field along one of the spin axes except for the spin-polarized phase.
On the other hand, the magnon gap is present in $\bm{h}\parallel b$.
We have confirmed the existence of the spin gap in the pure Kitaev model with $\alpha=-\pi/2$ in the magnetic field.
This result contrasts with that in the Majorana fermion picture in the Kitaev model under the effective magnetic field where the Majorana gap disappears when the magnetic field is applied along the $b$ direction~\cite{kitaev, arxiv2007-06757}.

\begin{figure*}[t]
  \begin{center}
    \includegraphics[width=180mm,clip]{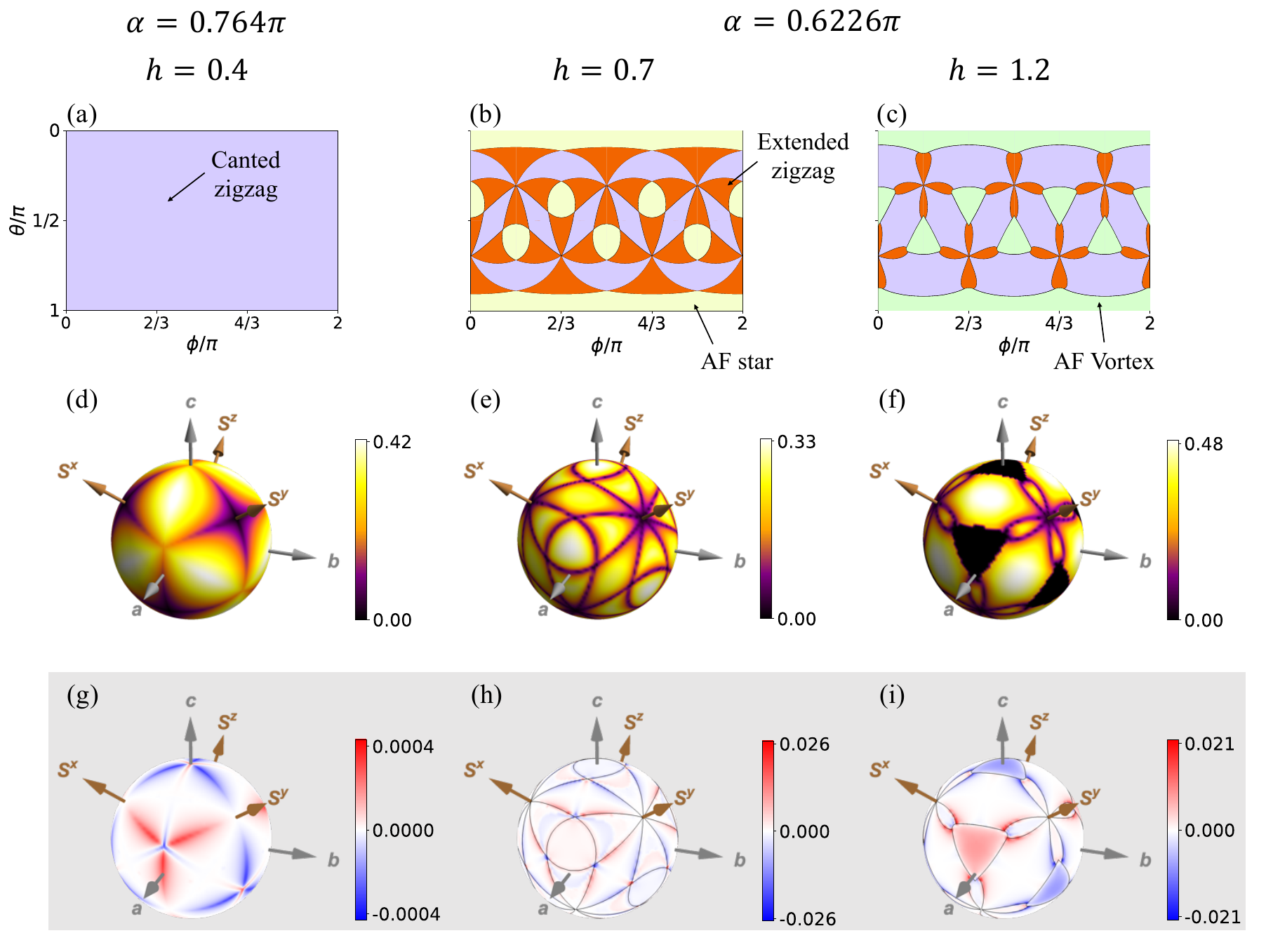}
    \caption{
(a) Classical phase diagram of the Kitaev-Heisenberg model with $\alpha = 0.764\pi$ on the plane of the field angles $(\theta,\phi)$ in the field strength fixed to $h=0.1$.
 (d),(g) Spherical plots of (d) the magnon gap and (g) $\kappa^{ab}/T$ at $T=0.05$ and $(\alpha, h) = (0.764\pi, 0.1)$.
(b),(e),(h) and (c),(f),(i) Corresponding plots at $(\alpha, h) = (0.6226\pi, 0.7)$ and $(0.6226\pi, 1.2)$, respectively.
   }
   \label{zigresult}
   \end{center}
\end{figure*}

Next, we show the field-angle dependence of the thermal Hall conductivity.
Figure~\ref{stresult}(g) shows the spherical plot of $\kappa^{ab}/T$ with respect to the field direction at $T=0.05$ and $h=0.4$.
The sign of $\kappa^{ab}$ changes on the spin axis planes similar to the case of the spin-polarized phase shown in Fig.~\ref{ffmth}(b).
The nodal lines along the spin axis planes appear commonly for the larger field strength [Fig.~\ref{stresult}(h) and \ref{stresult}(i)], and the existence of them is consistent with the symmetry analysis, which was discussed in Sec.~\ref{sec:symmetry}.
We find that the absolute value of $\kappa^{ab}$ is relatively small near the lines where the magnon gap takes minimal values.
On the other hand, a distinctly different feature is found in the angle dependence of $\kappa^{ab}$ for $h=0.63$ and $0.75$ around the $c$ axis.
As presented in Fig.~\ref{stresult}(h), $\kappa^{ab}$ is substantially enhanced in the vicinity of this direction for $h=0.63$, where the FM star phase appears.
This behavior is observed also for $h=0.75$, as shown in Fig.~\ref{ffmth}(i).
The results imply that the noncoplanar spin configurations play a crucial role in enhancing $\kappa^{ab}$ rather than the small magnon gap.
Nevertheless, $\kappa^{ab}$ in the FM star phase appears to be smaller than that in the canted stripy phase surrounding the FM star one.
This suggests that the magnitude of the thermal Hall conductivity is not simply determined by whether the spin configuration is coplanar or not.

\subsection{Canted zigzag, AF star, and AF vortex states}

\label{sec:zigzag}

\begin{figure}[t]
  \begin{center}
    \includegraphics[width=\columnwidth,clip]{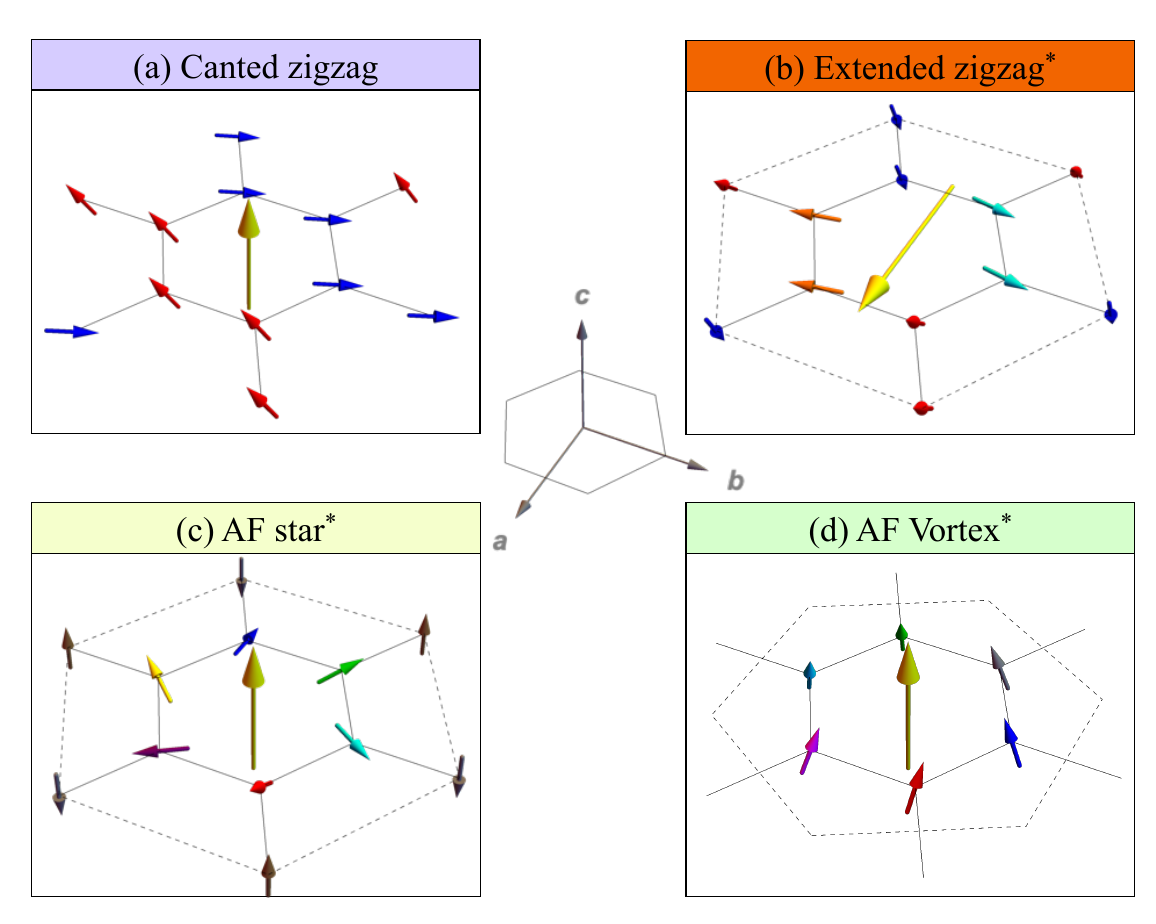}
    \caption{
 Spin configurations of (a) the canted zigzag, (b) extended zigzag, (c) AF star, and (d) AF vortex states.
 The magnetic unit cells are shown as dashed lines in (b)--(d).
 Non-coplanar spin configurations are indicated by asterisks.
 The yellow arrows represent the magnetic fields.
 Note that the field in (b) is applied along the $a$ direction.
    }
   \label{zigconfig}
   \end{center}
 \end{figure}

In this section, we show the cases where the zigzag state is stabilized in the absence of the magnetic field.
This state appears when $\pi/2< \alpha\lesssim 0.85\pi$~\cite{prl110-097204,khmodel}, where the Kitaev and Heisenberg interactions are antiferromagnetic (AF) and FM, respectively.
First, we focus on the case with $\alpha=0.764\pi$.
At this parameter, the canted zigzag and spin-polarized states are only induced by the magnetic fields along the $c$ and $S^z$ directions~\cite{khmodel}.
Figure~\ref{zigresult}(a) shows the phase diagram on the plane of the field angles $(\theta,\phi)$ at $(\alpha,\ h) = (0.764\pi,0.3)$.
We find that the canted zigzag state shown in Fig.~\ref{zigconfig}(a) appears regardless of the field angle.
Figure~\ref{zigresult}(d) shows the spherical plot of the magnon gap for the field angle.
The gap takes a minimal value on the lines where the field component taking the smallest value among $h_x$, $h_y$, and $h_z$, is switched.
This behavior is similar to the case of the canted stripy state presented in Fig.~\ref{stresult}(d).
On the other hand, the field-angle dependence of $\kappa^{ab}$ presented in Fig.~\ref{zigresult}(g) is distinctly different from that in Fig.~\ref{stresult}(g).
The thermal Hall conductivity takes a relatively large value on the lines from the $c$ axis to the spin axes, where the magnon gap is minimal.

Next, we show the results for $\alpha=0.6226\pi$.
In the case with $\pi/2< \alpha< 3\pi/4$, the zigzag state is destabilized, and the AF star state presented in Fig.~\ref{zigconfig}(c) appears immediately by the magnetic field parallel to the $c$ axis~\cite{khmodel}.
Figure~\ref{zigresult}(b) is the phase diagram on the plane of the field angles $(\theta, \phi)$ at $h=0.7$.
We find that the canted zigzag and extended zigzag states appear in addition to the AF star state.
Figure~\ref{zigconfig}(b) shows the extended zigzag state, which is a noncoplanar magnetic order, unlike the canted zigzag state shown in Fig.~\ref{zigconfig}(a). 
The AF star state appears in the vicinity of the $c$ direction, which is an eight-sublattice magnetic order with a noncoplanar spin configuration.
Figure~\ref{zigresult}(e) shows the spherical plot of the magnon gap at $(\alpha,h)=(0.6226\pi, 0.7)$.
The three phases appearing in Fig.~\ref{zigresult}(b) possess magnon gaps, but the gap disappears at the boundaries between them.
In particular, at $\bm{h}\parallel S^z$ and its equivalent directions, phase boundaries between canted zigzag and extended zigzag states are crossed, and the magnon gap vanishes.
We find that the thermal Hall conductivity is enhanced along the phase boundaries, presumably due to the suppression of the magnon gap as presented in Fig.~\ref{zigresult}(h).

With increasing the magnetic field along the $c$ direction, the AF vortex state is stabilized before entering the spin-polarized phase~\cite{khmodel}. 
This state is a six-sublattice magnetic order with a noncoplanar spin configuration as presented in Fig.~\ref{zigconfig}(d).
Figure~\ref{zigresult}(c) shows the phase diagram with respect to the field angles at $(\alpha,h)=(0.6226\pi, 1.2)$.
In addition to the AF vortex phase, the canted zigzag and extended zigzag phases are present in the phase diagram.
We find that the canted zigzag state is stable in the wider region than that at $h=0.7$, suggesting that an applied magnetic field stabilizes the canted zigzag state rather than the extended zigzag state.
In Figs.~\ref{zigresult}(f) and \ref{zigresult}(i), we show the spherical plots of the magnon gap and $\kappa^{ab}/T$ at $T=0.05$ for the field direction.
At the boundary between the canted zigzag and extended zigzag phases, the magnon gap vanishes.
In the vicinity of the boundary, the thermal Hall conductivity is enhanced as shown in Fig.~\ref{zigresult}(i).
These behaviors are similar to the case at $h=0.7$ [Figs.~\ref{zigresult}(e) and \ref{zigresult}(h)].
We also find that a gapless magnon mode is maintained in the AF vortex state.
In this state, $|\kappa^{ab}|$ takes a relatively large value [Fig.~\ref{zigresult}(i)] in comparison with that in the AF star phase shown in Fig.~\ref{zigresult}(h).
This feature is thought to originate from the gapless magnon mode with nonzero Berry curvature.
Note that, when the magnetic field is on the spin axis planes, the thermal Hall conductivity vanishes regardless of the field strength and the ordering pattern.

\begin{figure}[t]
  \begin{center}
    \includegraphics[width=\columnwidth,clip]{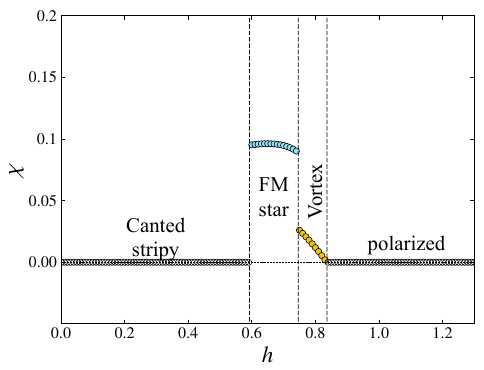}
    \caption{
(a) Scalar spin chirality $\chi$ as a function of the field strength at $\alpha = -0.312\pi$ for the field along the $c$ direction. 
   }
   \label{stcop}
   \end{center}
\end{figure}

\section{Discussion}
\label{sec:discussion}

In this section, we discuss the origin of the thermal Hall conductivity and the effect of spin interactions beyond the Kitaev-Heisenberg model.
As pointed out by the previous studies~\cite{jps86-011007, prl103-047203, cm29-3, prl104-066403}, spin configurations with the scalar spin chirality $\chi_{ijk}=\bm{S}_i \cdot (\bm{S}_j \times \bm{S}_k)$ defined for the neighboring three sites $[ ijk ]$ plays a crucial role in the appearance of the Hall effect in insulating magnets.
Here, we calculate the absolute value of the scalar spin chirality defined by $\chi=N_t^{-1}\sum_{[ ijk ]}|\braket{\bm{S}_i} \cdot \braket{\bm{S}_j} \times \braket{\bm{S}_k}|$, where $N_t$ is the number of neighboring three sites.
Figure~\ref{stcop} shows the field-strength dependence of $\chi$ in the Kitaev-Heisenberg model with $\alpha=-0.312\pi$ under the magnetic field applied along the $c$ direction.
As mentioned in Sec.~\ref{sec::st}, the FM star and vortex states appear, and the large thermal Hall conductivity is observed in these states as shown in Fig.~\ref{stmag}(b).
Since the FM star and vortex states exhibit non-coplanar spin configurations, $\chi$ is nonzero while it vanishes in the canted stripy and polarized phases.
The results suggest that spin configurations with the nonzero scalar spin chirality enhance the thermal Hall conductivity similar to conventional spin systems with the Dzyaloshinskii-Moriya interaction.
Note that, in the present case, the existence of the non-coplanar spin configurations is not a necessary condition for the nonzero thermal Hall conductivity; it also appears in coplanar spin configurations while its absolute value is small.

\begin{figure}[t]
  \begin{center}
    \includegraphics[width=\columnwidth,clip]{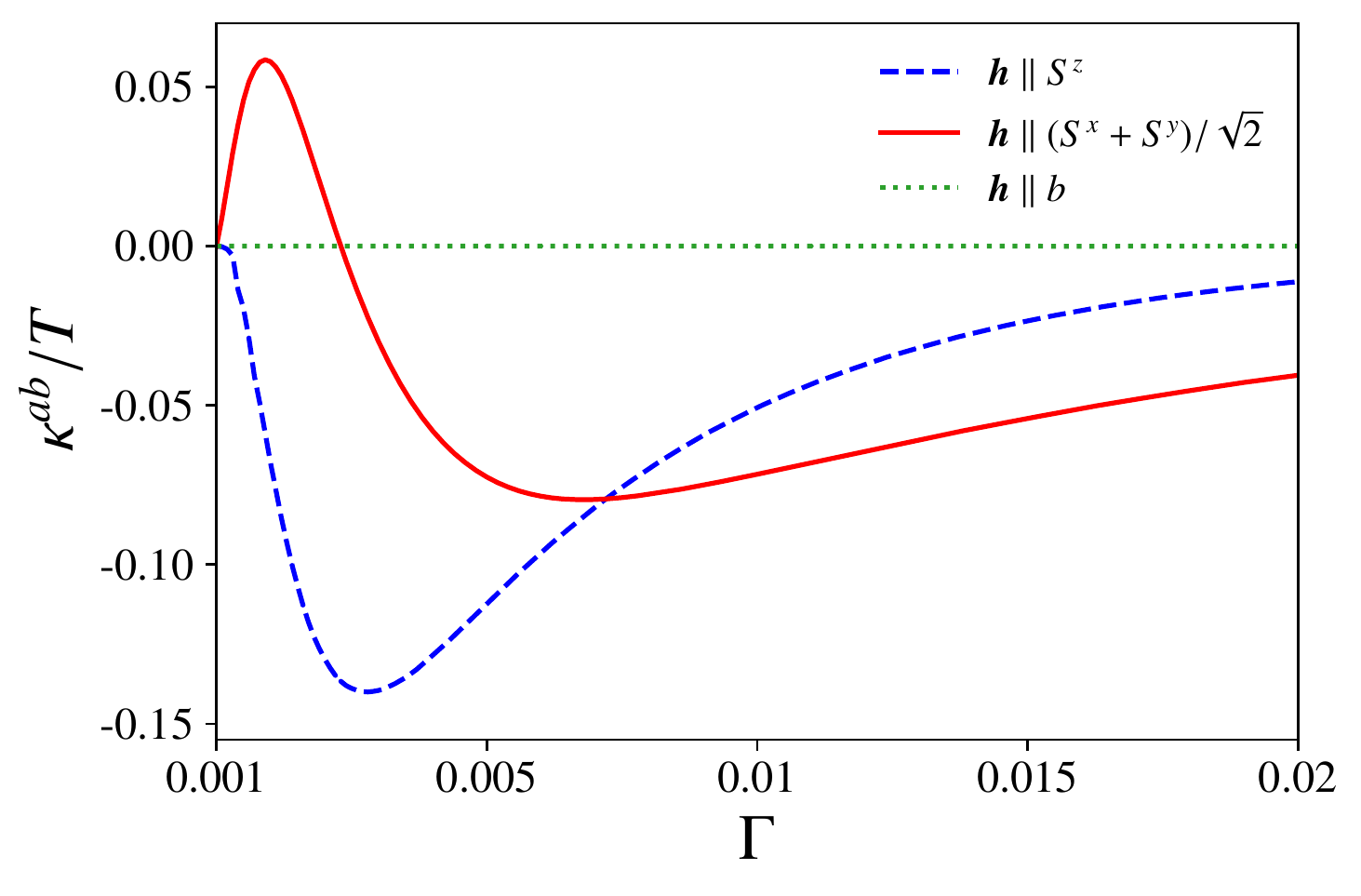}
    \caption{
Thermal Hall conductivity as a function of $\Gamma$ for the Kitaev-$\Gamma$ model at $(\alpha, h,T)=(-\pi/2,0.01, 0.05)$ with $\bm{h}\parallel S^z,\ (S^x+S^y)/\sqrt{2}$, and $b$.
   }
   \label{fig_gamma}
   \end{center}
\end{figure}

Finally, we comment on the effect of additional spin interactions to Eq.~\eqref{kh}.
As discussed in Sec.~\ref{sec:symmetry}, the thermal Hall conductivity vanishes in the Kitaev-Heisenberg model when the applied magnetic field is on the spin axis plane.
This feature originates from the fact that the Hamiltonian is represented as the real matrix because the neighboring spins interact with the same component.
It remains intact even in the presence of the long-ranged Heisenberg interactions~\cite{prb84-180407, prb93-214431}.
On the other hand, when off-diagonal spin interactions are introduced, the feature disappears, Here, we consider the $\Gamma$ interaction, which is one of the off-diagonal symmetric interactions~\cite{prl112-077204,nasu2014}.
This is given by
\begin{align}
  {\cal H}_{\Gamma} = \Gamma \sum_{\langle ij\rangle_{\gamma}} \left(S_i^{\alpha} S_j^{\beta}+S_i^{\beta} S_j^{\alpha}\right),
\end{align}
where $(\alpha,\beta,\gamma)=(x,y,z)$ and its cyclic permutations.
To clarify the effect of the $\Gamma$ interaction, we calculate the thermal Hall conductivity in the Kitaev-$\Gamma$ model with the magnetic field applied along $S^z$, $(S^x+S^y)/\sqrt{2}$, and the $b$ axis.
In the absence of the $\Gamma$ interaction, $\kappa^{ab}$ is zero as these fields are on the spin axis plane.
Figure~\ref{fig_gamma} shows the $\Gamma$ dependence of the thermal Hall conductivity.
As presented in this figure, $\kappa^{ab}$ becomes nonzero for $\bm{h}\parallel S^z$ and $(S^x+S^y)/\sqrt{2}$ by introducing the $\Gamma$ interaction.
The thermal Hall conductivity shows the nonmonotonic $\Gamma$ dependence and appears to approach zero with increasing $\Gamma$.
On the other hand, $\kappa^{ab}$ remains zero for $\bm{h}\parallel b$.
We also note that the above discussion is based on the properties of the Hamiltonian, and it is applicable even in the case beyond the spin-wave approximation.
Therefore, we expect that the magnitude of off-diagonal interactions such as the $\Gamma$ interaction can be generally deduced from the field-angle dependence of the thermal Hall conductivity in experiments.
In the measurement of $\kappa^{ab}$ obtained by changing the field angle from $c$ to $a$, the deviation of the field angle satisfying $\kappa^{ab}=0$ from the $(S^x+S^y)/\sqrt{2}$ direction is interpreted as a consequence of the off-diagonal interactions.

\section{Summary}
\label{sec:summary}

In summary, we have revealed the magnetic-field effect on the thermal transport in the Kitaev-Heisenberg model using the mean-field approximation and the spin-wave theory.
We have demonstrated that the field-angle dependence of the thermal Hall conductivity caused by the magnon excitations strongly depends on the spin structure of the magnetic order.
In particular, noncoplanar spin configurations with nonzero spin scalar chirality enhance the thermal Hall conductivity.
We have also found that it shows common nodal lines on the sphere describing the field angle.
The lines are located on the spin axis planes, which is an intrinsic feature of the Kitaev-Heisenberg model.
The feature disappears by introducing the off-diagonal spin interactions such as the $\Gamma$ terms.
Therefore, the present results not only provide comprehensive data on the thermal Hall conductivity under the variation of the field direction but also shed light on the role of the off-diagonal interactions.
This will stimulate further experimental studies on the transport phenomena in Kitaev-related systems.

\begin{acknowledgments}
The authors thank S.~Hayami for fruitful discussions.
Parts of the numerical calculations were performed in the supercomputing
systems in ISSP, the University of Tokyo.
This work was supported by Grant-in-Aid for Scientific Research from JSPS, KAKENHI Grant No.~JP19K03742, JP20H00122, and JST PREST Grant No.~JPMJPR19L5 (J.N.).
\end{acknowledgments}

\bibliography{./refs}
\end{document}